# Heterogeneous Influence Maximization in User Recommendation

Hongru Hou*
School of Data Science, Fudan
University
Shanghai, China
hongruhou89@gmail.com

Jiachen Sun
Tencent
Shenzhen, China
jiachensun@tencent.com

Wenqing Lin†
Tencent
Shenzhen, China
edwlin@me.com

Wendong Bi
Tencent
Shenzhen, China
wendongbi@tencent.com

Xiangrong Wang†
Shenzhen University
Shenzhen, China
xiangrongwang88@gmail.com

Deqing Yang
School of Data Science, Fudan
University
Shanghai, China
yangdeqing@fudan.edu.cn

## Abstract

User recommendation systems enhance user engagement by encouraging users to act as inviters to interact with other users (invitees), potentially fostering information propagation. Conventional recommendation methods typically focus on modeling interaction willingness. Influence-Maximization (IM) methods focus on identifying a set of users to maximize the information propagation. However, existing methods face two significant challenges. First, recommendation methods fail to unleash the candidates' spread capability. Second, IM methods fail to account for the willingness to interact. To solve these issues, we propose two models named HeteroIR and HeteroIM. HeteroIR provides an intuitive solution to unleash the dissemination potential of user recommendation systems. HeteroIM fills the gap between the IM method and the recommendation task, improving interaction willingness and maximizing spread coverage. The HeteroIR introduces a two-stage framework to estimate the spread profits. The HeteroIM incrementally selects the most influential invitee to recommend and rerank based on the number of reverse reachable (RR) sets containing inviters and invitees. RR set denotes a set of nodes that can reach a target via propagation. Extensive experiments show that HeteroIR and HeteroIM significantly outperform the state-of-the-art baselines with the p-value<0.05. Furthermore, we have deployed HeteroIR and HeteroIM in Tencent's online gaming platforms and gained an 8.5% and 10% improvement in the online A/B test, respectively. Implementation codes are available at https://github.com/socialalgo/HIM.

*This work was done while Hongru Hou was an intern at Tencent.
†Corresponding author

## 1 Introduction

Recommendation systems operated by providing a list of candidates to users are widely deployed, from social media [2, 16, 32] to e-commerce [18, 23, 36] and online gaming platforms [25, 26, 37, 40–42]. User recommendation systems, as one of the existing manners, primarily focus on encouraging more users (inviters) to invite other users (invitees) to increase overall engagement. For instance, team recommendations in online gaming platforms, which suggest potential friends for collaborative gameplay, can significantly enhance user engagement and retention [20, 22].

In addition to the aforementioned benefits, user recommendation systems can significantly foster the propagation of information [16, 29]. To promote up-to-date gameplay mechanics, online gaming platforms frequently employ both external advertisements [13] and in-game events with incentives [17]. External advertising, however, is often costly and may yield insufficient returns on investment [1]. In contrast, in-game activities that motivate users to invite friends can effectively propagate information across social networks, as users are more inclined to accept information from acquaintances [3, 5, 11]. In in-game activities, recommendation systems assist users in deciding whom to invite, facilitating easier participation for users and improving engagement. As a result, the recommendation tailored to user preferences can achieve broader dissemination without incurring additional advertising costs.

To make the recommendations align with user preferences, extensive click-through rate (CTR) models have been developed [12, 31, 35, 43, 44]. These methods mainly focus on designing different feature interaction methods to improve prediction accuracy. For instance, AutoInt [31] employs an attention mechanism to learn and weigh the importance of different features dynamically, and Eulernet [35] learns feature interactions in a complex vector space by conducting space mapping according to Euler's formula. All these recommendation models perform well in scenarios where only the user's click action is required. However, in user recommendation, a broader information propagation is significant. The aforementioned models fail to unleash the spread potential. Specifically, invitees have the potential to become inviters themselves and subsequently dominate further invitation activities, leading to broader dissemination.

To fully leverage the capabilities of user recommendation systems in information dissemination, it is essential for the recommendation to consider the invitees' spread capability. Zhang et al. [40] propose a model RR-OPIM+, which incrementally selects invitees with the highest spread capability and recommends them to inviters. While this approach can recommend the invitees with the highest spread capability, it fails to take into account the inviters' willingness to interact, which may lead to ineffective recommendations. Indeed, the solution to maximize the spread coverage while maintaining interaction willingness remains a significant obstacle, highlighting a critical area for improvement.

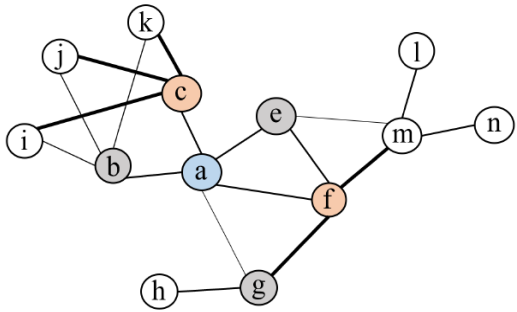

**Figure 1: Schematic presentation of the optimal recommended list of user $a$. Recommending users $c$ and $f$ to user $a$ fosters the largest spreading coverage if each user only shares information with up to two friends. The thickness of the edges signifies the likelihood of interaction occurring.**

To overcome such a challenge, it is essential to account for spread capability and interaction willingness simultaneously. As illustrated in Figure 1, we present the optimal recommendation candidates (orange) for user $a$, which includes users $c$ and $f$. The reasons are as follows: (i) user $a$ has the potential to interact with user $c$ and $f$, (ii) users $c$ and $f$ possess strong spread capability, and (iii) users $c$ and $f$ can spread to different groups of people which can maximize the secondary spread coverage in global.

**Contributions**. In light of the aforementioned limitations, we aim to empower the conventional recommendation methods with spread potential and bridge the gap between the IM methods and the recommendation task. To accomplish this, we propose two models called HeteroIR (Heterogeneous Influence-based Recommendation) and HeteroIM (Heterogeneous Influence-Maximization Recommendation), which achieve the goals above, respectively. The core idea of HeteroIR is to integrate the spread capability into the quantification of recommended profits, while HeteroIM takes the heterogeneous interaction willingness into the ranking of the candidates. Our new recommendation algorithms aim to introduce a broader information propagation while preserving the accuracy and efficiency of the recommendations. To begin, we present a spread influence algorithm, HeteroInf (Heterogeneous Influence), to estimate the spread capability. Furthermore, we introduce a model-agnostic framework, HeteroIR, to quantify the spread profits of the recommendations given. Moreover, we develop an IM framework empowered by shared RR sets to quantify the likelihood of each user pair interacting, which fulfills the recommendation with both high spread capacity and interaction willingness.

We experimentally evaluate the HeteroIR and HeteroIM proposed against seven representative competitors on three datasets. The results show our algorithms outperform the competitors in terms of the spread metric NSpread@K and the recommendation metric Recall@K and NDCG@K. Furthermore, we deploy our solutions in two real-world scenarios on Tencent's online gaming platforms. Here, we estimate the number of invitees in secondary spread times, secondary spread ratio, and retain ratio. Compared with the baseline model, relative improvements of up to 10%, 9.64%, and 14.83% are achieved in corresponding evaluation metrics, respectively.

To summarize, we make the following contributions in this work:

- We propose HeteroIR, an intuitive recommendation algorithm that effectively integrates the spread influence with interaction willingness.
- We introduce HeteroIM, a recommendation algorithm grounded in influence maximization, designed to fill the gap between influence maximization and recommendation tasks.
- We validate the performance of the HeteroIR and HeteroIM in three datasets, which outperforms the state-of-the-art baselines in both the spread task (NSpread@K) and the recommendation tasks (Recall@K and NDCG@K).
- We have deployed the HeteroIR and HeteroIM to two user recommendation events in Tencent's online games, achieving significant improvements compared with the baseline model.

## 2 PRELIMINARIES

This section introduces the in-game user recommendation task and the problem formulated in the work.

### 2.1 In-Game User Recommendation

On Tencent's online gaming platforms, the service provider regularly organizes invitation-based activities to foster user interactions and enhance user engagement. Before an event, the service provider selects a set of users $\mathcal{V}_i$ as the inviters and a set of users $\mathcal{V}_e$ as the invitees. For each inviter $u \in \mathcal{V}_i$, a limited number of friends are selected in terms of specific recommendation algorithms. As the event commences, each inviter receives the event details and a recommendation list that includes invitees encouraged to interact. Since receiving the invitation from the inviter $u$, the invitee $v$ is notified and has a chance to decide whether to accept the invitation. Once accepted, a valid recommendation from $u$ to $v$ is conceived.

### 2.2 PROBLEM FORMULATION

In a directed and weighted attributed Graph $G = (V, E, \mathbf{P}, \mathbf{U})$, let $V = \{v_1, v_2, ..., v_N\}$ be a set of nodes and $E$ be a set of edges. $\mathbf{P} \in \mathbb{R}^{N \times N}$ represents the weighted adjacency matrix where $P_{uv}$ denotes the probability user $u$ invites user $v$ and $\mathbf{U} \in \mathbb{R}^N$ represents the node attribute vector where $U_j$ denotes the probability user $j$ accept the invitation from any inviters.

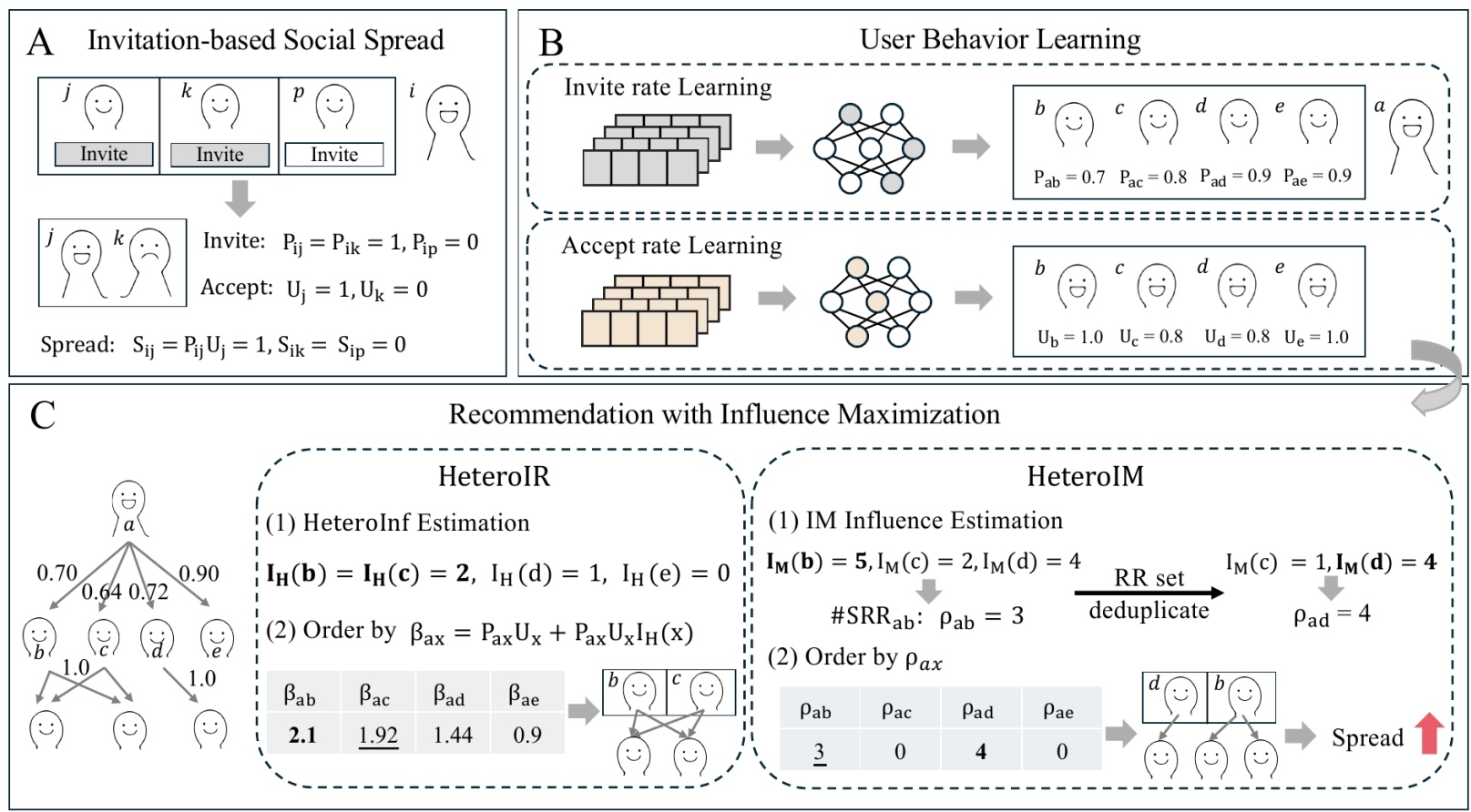

**Figure 2: The overall framework of our proposed methods. Panel (A) illustrates the diffusion process in user recommendation, which consists of the invite and accept stages. Panel (B) shows the learning of user behaviors, where the invite rate and accept rate are modeled independently. Panel (C) depicts the details of the HeteroIR and HeteroIM. In HeteroIR, we first estimate the spread influence by HeteroInf, which is further extended to the ranking function $\beta$. For HeteroIM, we estimate the spread influence based on the RR sets and further rerank the candidates based on the number of Shared RR sets (SRR) $\rho$.**

PROBLEM 1 (RECOMMENDATION WITH INFLUENCE MAXIMIZATION (RIM)). *The task of* RIM *involves giving a capacity-limited recommend list to gain a broader spread coverage while maintaining interaction willingness in attributed graph $G = (V, E, \mathbf{P}, \mathbf{U})$.*

User recommendation plays a pivotal role in information propagation as information can propagate on the social network through interactions among users. These activities foster interactions and enable the exchange of information among users, enabling those who are unfamiliar with the information to be informed. Besides the direct interactions between inviters and invitees, invitees also possess the potential to become inviters themselves, motivated by incentives. The invitees' potential in dissemination generates a ripple effect, enabling information to propagate swiftly throughout the social network and ultimately culminating in a viral marketing campaign. By optimizing the recommendation algorithms, we can propagate information more efficiently while maintaining the effectiveness of the recommendations given.

## 3 FRAMEWORK

In this section, we introduce two methods to cope with the RIM problem, and the framework is shown in Figure 2. Section 4.1 briefly illustrates the diffusion model under the user recommendation and corresponding spread probability modeling. Section 4.2 delves into the recommendation algorithm that integrates spread influence. Section 4.3 elaborates on the recommendation with influence maximization.

### 3.1 Spread Probability Modeling

**Diffusion Model**. In game-social scenarios, the completion of a spread involves both an invitation from the inviter and the invitee's acceptance, as shown in Figure 2A. We abstract this two-stage task as a diffusion model as follows:

(1) Inviter $u$ sends an invitation to friend $v$ with probability $P_{uv}$.

(2) Invitee $v$ accepts an invitation from any inviters with $U_v$.

(3) The probability of social diffusion happens $S_{uv}$ among $(u, v)$ is characterized by $P_{uv}U_v$.

**User Behavior Learning**. By collecting the inviting and accepting data from the dataset, two models are trained for the prediction of invite probability $P_{uv}$ and accept probability $U_v$ respectively. The heterogeneous spread probability $S_{uv}$ among the user pair $(u, v)$ equals $P_{uv}U_v$.

### 3.2 Recommend With Spread Influence

Based on the diffusion model and spread probability predicted above, we can estimate the spreading influence $I(u)$ of each user $u$ in the network. We first define the spread influence as follows:

DEFINITION 1 (SPREAD INFLUENCE $I$). *Given a social network $G = (V, E)$. For any user $u \in V$, the spread influence $I(u)$ is defined as the expected number of users influenced by user $u$.*

We collect user logs from two incentive propagation events in Tencent's first-person shooter game. Specifically, we find that each user has 50 friends on average, while the spread influence on average is 16 times smaller. Moreover, we observe that the user's friend number and spread influence share a low correlation, with a Pearson correlation [6] equal to 0.06.

OBSERVATION 1 (LOW SPREAD INFLUENCE). *The spread influence of the user in the game social network is far less than the number of friends, while sharing a low correlation.*

The observation above indicates that the spread capability is limited compared to the friend number in game social networks where the methods [19, 38, 39] fully considering the local structure, overestimate the spread influence in real scenarios.

**Heterogeneous Influence Algorithm.** Motivated by this observation, we design a heterogeneous influence algorithm, HeteroInf, to evaluate the spread influence of each user with interaction capability $w$ limited. Firstly, we select a set of out-neighbors $N_u^w$ of user $u$ with the spreading probability $S_{u,v}$ ranked top $w$ among all the neighbors. Secondly, the influence of user $u$ is characterized by the sum of the spread probability from $u$ to the user in $N_u^w$. The HeteroInf for the node $u$ aggregating $w$ neighbors is formulated as Equation 1:

$$I_H(u) = \sum_{v \in N_u^w} P_{uv} U_v \quad (1)$$

where $P_{uv}$ denotes the invite probability and $U_v$ denotes the accept probability of user $v$. $w$ is the interaction capability.

**Heterogeneous Influence-based Recommendation.** By taking the spread influence of the invitees and the personalized behavior from both parties, we introduce an algorithm named HeteroIR. The ranking function $\beta_{uv}$ is shown in Equation 2. This equation illustrates the profits of recommending candidate $v$ to user $u$, which consists of two components: $P_{uv}U_v$ represents the probability of first-order influence from $u$ to $v$ happens, while $P_{uv}U_v I(v)$ captures the profit of secondary influence facilitated by candidate $v$. $I_H(v)$ denotes the spread influence of user $v$ estimated by HeteroInf.

$$\beta_{uv} = \underbrace{P_{uv}U_v}_{\text{1st-IF}} + \underbrace{P_{uv}U_v I_H(v)}_{\text{2nd-IF}} \quad (2)$$

## 3.3 Recommend With Influence Maximization

Despite HeteroIR's incorporation of spreading influence, it overlooks the issue of spreading overlap among the suggested invitees, as the invitees $b$ and $c$ recommended spread to the same person shown in Figure 2C. Moreover, existing IM methods fail to consider the interaction willingness. To further maximize the spread coverage while maintaining interaction willingness, we propose a heterogeneous influence-maximization recommendation, HeteroIM, making the IM framework cater to the demands of recommendation tasks. Prior to HeteroIM, we first define the Reverse reachable set (RR set) proposed by Borgs et.al [4].

DEFINITION 2 (RR SET). *Given a graph $G = (V, E)$ and a diffusion model $M$, a random RR set $R_{G,M}$ is a set of nodes, generated by (i) randomly selecting $v \in V$ as source node; (ii) reversly sampling the set $R_{G,M}$ of nodes that can spread to $v$ in terms of $M$.*

In the maximization framework based on the RR set, the spread capability of user $u$ can be gauged by the number of RR sets covering $u$ [40]. Firstly, we generate a number of RR sets $R_{G,M}$ with

**Algorithm 1:** HeteroIM $(A, N, R_{G,M}, k)$

1 **while** $N \neq \emptyset$ **do**
2      calculate the covered times of each node $c_i$ by $R_{G,M}$;
3      $u \leftarrow \arg\max_{u \in N} c_u$ ▷ select the most influential user;
4      **foreach** *neighbor $v$ of $u$* **do**
5          $L[v] \leftarrow u$ ▷ recommend $u$ to the user $v$;
6          $S[v] \leftarrow [\rho_{uv}, c_u]$ ▷ $\rho_{uv}$: # SRR of $u$ and $v$;
7      $N \leftarrow N \setminus u$; $R_{G,M} \leftarrow R_{G,M} \setminus R'_{G,M}$;
8 **foreach** $u \in A$ **do**
9      sort $L[u]$ by $S[u][0]$ then $S[u][1]$ descending ▷ <u>Rerank</u>;
10      select top $k$ user in $L[u]$ as the recommended of user $u$;

IC model [10] and **heterogeneous spread probability (HSP)** $S_{uv}$ elaborated in Section 4.1. HSP considers the personalized interaction probability instead of homogeneous probability [4, 40]. The RR set number generated $RN$ follows the RR-OPIM+ [40] as shown in Equation 3:

$$RN = 2^{i_{max}} \cdot \theta, \quad (3)$$

where

$$i_{max} = \left\lceil \log_2 \frac{n_p}{k \cdot \chi \cdot \epsilon^2} \right\rceil \quad (4)$$

$$\theta = 2 \cdot \left( \frac{1}{2}\sqrt{\ln \frac{6}{\delta}} + \sqrt{\frac{1}{2} \cdot \left(\ln\left(\prod_{u \in A} \binom{|C_u|}{k}\right) + \ln \frac{6}{\delta}\right)} \right)^2 \quad (5)$$

$A$ denotes the inviter set, $n_p$ denotes #nodes in $G$, $k$ denotes recommendation length, $\delta, \epsilon$ denotes error constant, $\chi$ denotes the size of inviter set, and $C_u$ denotes the out-degree of the user $u$.

DEFINITION 3 (SHARED RR SET). *Given a set of RR sets $R_{G,M}$, the subset $R'_{G,M} \subseteq R_{G,M}$ containing both node $i$ and node $j$ is denoted as the shared RR sets (SRR) of node $i$ and node $j$.*

To avoid sampling bias introduced by source nodes. We utilize a **uniform sampling (US)** strategy. Specifically, we select each node as the source node to perform RR sets generation for $\lfloor \frac{RN}{N} \rfloor$ times. Moreover, we select the first-order neighbor of $A$ to get the candidate set $N$. Subsequently, we take the RR sets $R_{G,M}$ as input for the following HeteroIM algorithm. HeteroIM can be divided into two steps: (*i*) Generate recommendation lists $L$ based on $R_{G,M}$; (*ii*) **Rerank** the recommendations based on the number of shared RR sets. The pseudocode for HeteroIM can be shown in Algorithm 1.

The graphic representation for the HeteroIM algorithm is depicted in Figure 2C. After the sample of $R_{G,M}$, user $b$ is the most influential candidate as it is covered by the most RR sets. Hence, we first recommend user $b$ to $a$ and remove the RR sets covering $b$ to decrease the spread overlap. Further, we recommend the user $d$ to user $a$ based on the RR sets remained.

To align the recommendations given by IM methods with interaction willingness, we rerank the recommendations based on the quantity of shared RR sets $\rho$. The user pair shares more RR sets, exhibiting a higher interaction probability. Hence, we first recommend user $d$ to user $a$, then recommend user $b$ to user $a$.

## 4 EXPERIMENTS

This section presents the experimental evaluation of HeteroIR and HeteroIM on multiple datasets to answer the following questions:

- **RQ1**: Do our proposed HeteroIR and HeteroIM improve upon existing state-of-the-art recommenders considering recommendation and IM methods across various experimental settings?
- **RQ2**: Are the key components in our HeteroIR and HeteroIM delivering the expected performance gains?
- **RQ3**: Does the Influence algorithm HeteroInf perform better compared with the existing influence methods?
- **RQ4**: Do HeteroIR and HeteroIM gain improvement when deployed online?

## 4.1 Experimental Setups

**Datasets**. We conduct evaluations of HeteroIR and HeteroIM on three datasets as shown in Table 1. **TXG**: This dataset contains user relationships in the Tencent game platform and corresponding invite and accept behavior among these relationships and users. **Twitter** [7]: This dataset is a widely used spread dataset which consists of user relationships in Twitter platforms and corresponding interaction behaviors among users, such as mention.

**Table 1: Dataset statistics ($M = 10^6$).**

| Dataset | #nodes (n) | #edges (m) | #spreads (s) |
|---|---|---|---|
| **TXG-A** | $109.1M$ | $177.1M$ | $62.4M$ |
| **TXG-B** | $131.6M$ | $211.4M$ | $43.2M$ |
| **TXG-C** | $115.5M$ | $179.4M$ | $35.2M$ |
| **Twitter** | $0.46M$ | $14.9M$ | $0.15M$ |

**Dataset Cleaning**. For TXG datasets, we collect logs from the friendship-centric social event in Tencent's first-person shooter game. The dataset is bifurcated into two parts: ($i$) An Exposure-Invitation dataset comprising tuples $(u, v, T_{u,v}, P_{u,v})$, which signifies that the invitee $v$ was exposed to the inviter $u$ at timestamp $T_{u,v}$ with an invitation from $u$ to $v$ issued if $P_{u,v} = 1$; ($ii$) An acceptance dataset containing tuples $(v, U_v)$, indicating that the target user $v$ received the invitation from one of the inviters end with acceptance if $U_v = 1$. For the Twitter dataset, we take the mentioned action as the spread process, once user $u$ mentions $v$ in a tweet, the spread from $u$ to $v$ is conceived.

**Spread Probability Modeling.** For TXG datasets, we trained two EulerNet [35] models on the TXG-A dataset to predict $P_{uv}$ and $U_v$ based on the user profile features. Further, we validated on the TXG-B and TXG-C datasets. We divided the TXG-A dataset into training, validation, and testing sets using an 8:1:1 ratio.

For the Twitter dataset, we first split the spreader into training and testing using an 8:2 ratio. Further, we sample the effective spread trajectory originating from the user in the training set as positive samples, the edges originating from the node in the training set without spread as negative samples, keeping a 1:1 ratio. Finally, we utilize EulerNet to fit the spread probability $S_{uv}$ based on the pre-trained network embedding, Deepwalk [27, 28], with 64-dim.

**Evaluation Protocols and Metrics.** To ensure a comprehensive evaluation, we conduct an evaluation on both the spread and recommendation tasks. For the spread task, we introduce a metric, Spread@K, which denotes the effective spread coverage for the given candidates. Specifically, this metric evaluates the effective first-order spread from the inviter to the top $k$ candidates, plus the spread coverage originating from the top $k$ candidates recommended. Noticeably, the spread coverage is deduplicated. To make the value comparable, we normalized the Spread@K by ISpread@K, which denotes the upper limit of the spread coverage given $k$ candidates, as NSpread@K. Here, we select $k$ friends generating the largest spread coverage for each user. The spread coverage originated from the sets comprising all these $k$ friends as the ISpread@K.

The graph representation for the calculation of NSpread@K is illustrated in Figure 3. For example, considering $K = 1$, the optimal recommended friends for users X and Y are X-#3 and Y-#3 as X-#3 and Y-#3 spread broader compared with other neighbors. The ISpread@1 first contains X-#3 and Y#3 as they are spread by X and Y successfully. Furthermore, X-#3 spreads to B, C, and D while Y-#3 spreads to F and G. Hence, ISpread@1 equals to 7. In the Tencent Games recommendation scenario, the average click times of a user with a click action is 3. Hence, we consider $K = 1, 2, 3$.

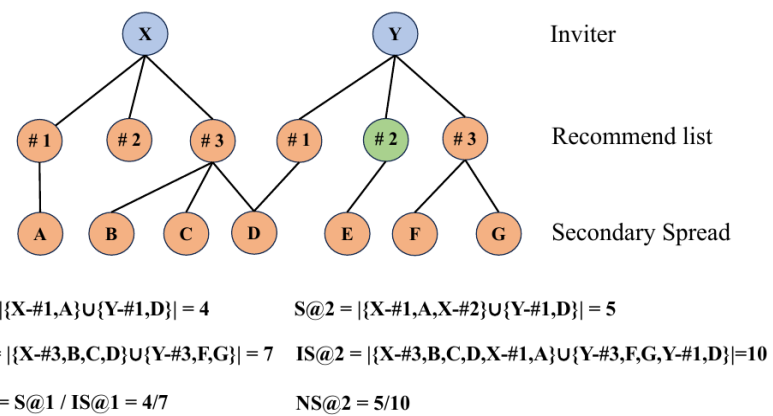

**Figure 3: Graphical representation for the calculation of NSpread@K. #$i$ denotes the invitee in the $i$-th exposure position. The orange denotes the invitee who was invited and accepted the invitation. The green denotes the invitee who was not invited or was being invited, but without acceptance, which should not be taken into account for Spread@K. The NSpread@K is a normalized value of the Spread@K that compares the actual secondary spread coverage of a recommendation list to the Ideal spread coverage ISpread@K.**

For the recommendation task, we use two widely adopted ranking-based metrics: Recall@N and NDCG@N, which measure the model's effectiveness in user ranking.

**Baselines**. We evaluate the effectiveness of our HeteroIR and HeteroIM in the spread and recommendation tasks with the state-of-the-art IM and recommendation methods. The IM methods focus on identifying a small number of influential users to maximize the information propagation, and we adopt widely-used IMM [34], OPIM-C [33], RR-OPIM+ [40] as the IM-based baselines. The recommendation methods focus on modeling the interaction probability between each user pair, and we adopt Personalized PageRank (PPR) [24], and supervised methods such as AutoInt [31], FinalNet [44], EulerNet [35] as recommendation baselines.

To validate the HeteroInf proposed, structural-based influence methods (Degree [38], Coreness [21], Windex [39]), simulation-based methods (MC influence [19]), and learning based methods (LR influence [9], TOPSIS [8], DeepInf [28]) are adopted as baselines.

For fair comparison, we set HeteroIM, OPIM-C [33], and RR-OPIM+ [40] with $\epsilon = 0.1$ and $\delta = 1/n$. We determine the hyperparameters $w$ for HeteroIR through grid search.

Table 2: Recommendation performance Improvement of all models on different datasets in terms of NSpread@K, Recall@K, and NDCG@K. The best performances are highlighted in bold, and the second-best are underlined. The superscript * indicates the Improvement is statistically significant where the p-value is less than $0.05$.

| Dataset | TXG-B | | | | | | TXG-C | | | | | | Twitter | | | | | |
|---|---|---|---|---|---|---|---|---|---|---|---|---|---|---|---|---|---|---|
| Model | NS@1 | NS@3 | R@1 | R@3 | N@1 | N@3 | NS@1 | NS@3 | R@1 | R@3 | N@1 | N@3 | NS@1 | NS@3 | R@1 | R@3 | N@1 | N@3 |
| IMM [34] | 0.3457 | 0.5254 | 0.1654 | 0.4599 | 0.2255 | 0.3008 | 0.3952 | 0.5711 | 0.2849 | 0.5805 | 0.3465 | 0.4232 | 0.1846 | 0.2876 | 0.0938 | 0.1833 | 0.1173 | 0.1537 |
| OPIM-C [33] | 0.3480 | 0.5269 | 0.1724 | 0.4653 | 0.2322 | 0.3087 | 0.3976 | 0.5732 | 0.2902 | 0.5861 | 0.3534 | 0.4302 | 0.1818 | 0.2955 | 0.0972 | 0.1906 | 0.1212 | 0.1594 |
| RR-OPIM+ [40] | 0.3520 | 0.5310 | 0.1798 | 0.4659 | 0.2399 | 0.3154 | 0.3999 | 0.5767 | 0.2977 | 0.5871 | 0.3606 | 0.4368 | 0.1792 | 0.2991 | 0.0946 | 0.1922 | 0.1188 | 0.1595 |
| PPR [24] | 0.3255 | 0.4688 | 0.1453 | 0.3193 | 0.1924 | 0.2798 | 0.3742 | 0.5134 | 0.2632 | 0.4396 | 0.3141 | 0.4022 | 0.1640 | 0.3372 | 0.1121 | 0.3395 | 0.1359 | 0.2520 |
| AutoInt [31] | 0.3879 | 0.5548 | 0.1922 | 0.4935 | 0.2500 | 0.3375 | 0.4369 | 0.6022 | 0.3102 | 0.6149 | 0.3711 | 0.4596 | 0.2234 | 0.4007 | 0.2721 | 0.4894 | 0.3140 | 0.4142 |
| FinalNet [44] | 0.3925 | 0.5579 | 0.1986 | 0.5037 | 0.2578 | 0.3487 | 0.4411 | 0.6054 | 0.3166 | 0.6260 | 0.3791 | 0.4704 | 0.2249 | 0.4071 | 0.2817 | 0.4985 | 0.3261 | 0.4244 |
| EulerNet [35] | 0.3949 | 0.5598 | 0.2021 | 0.5117 | 0.2619 | 0.3532 | 0.4435 | 0.6081 | 0.3202 | 0.6325 | 0.3832 | 0.4758 | 0.2258 | 0.3973 | 0.2849 | 0.4944 | 0.3302 | 0.4237 |
| HeteroIR | 0.4253* | 0.5813* | 0.2278* | 0.5254* | 0.2959* | 0.3822* | 0.4712* | 0.6295* | 0.3466* | 0.6477* | 0.4173* | 0.5034* | 0.2432* | 0.4185* | 0.2933* | 0.5095* | 0.3402* | 0.4375* |
| **HeteroIM** | **0.4398*** | **0.5979*** | **0.2303*** | **0.5478*** | **0.2998*** | **0.3878*** | **0.4849*** | **0.6443*** | **0.3505*** | **0.6701*** | **0.4204*** | **0.5065*** | **0.2445*** | **0.4222*** | **0.3655*** | **0.5721*** | **0.4284*** | **0.5102*** |
| Best Imprv. | ↑11.39% | ↑6.79% | ↑13.96% | ↑7.06% | ↑14.50% | ↑9.80% | ↑9.33% | ↑5.94% | ↑9.47% | ↑5.94% | ↑9.71% | ↑6.45% | ↑8.28% | ↑3.71% | ↑28.29% | ↑14.76% | ↑29.73% | ↑20.22% |

## 4.2 Performance Comparison (RQ1)

To validate the proposed recommendation models, we compare the performance of each model in spread and recommendation task.
**Spread Evaluation.** Table 2 shows the NSpread@K of different algorithms. The recommendations given by the HeteroIR spread broader than the probability given by recommendation models since the HeteroIR takes the spreading capability of candidates into consideration. By leveraging the spreading overlap among candidates, HeteroIM gets the best performance, consistently showing the spread overlap among the candidates.
**Recommendation Evaluation.** For the evaluation of the recommendation task, Recall@K and NDCG@K were used. In TXG datasets, we regard interaction undergone both the inviter's click and the invitee's acceptance as the ground truth for the calculation of Recall@K and NDCG@K. For the Twitter datasets, the mention interaction serves as a valid recommendation. The results are presented in Table 2. Our model demonstrates superior performance, suggesting that HeteroIR and HeteroIM leverage both enhanced spreading capability and improved performance on recommendation tasks simultaneously.

## 4.3 Ablation Study (RQ2)

To study the impact of the main components of HeteroIR and HeteroIM, we conduct ablation studies as follows.
**HeteroIR.** In HeteroIR, we consider both first-order (1st-IF) and second-order influence (2nd-IF) to calculate the spread profits. The result of the ablation study on these two parts is shown in Table 3. We observe that by removing the 1st-IF, the Recall and NDCG decrease significantly, as 1st-IF considers the interaction willingness of the recommendation. Moreover, by removing the 2nd-IF, the Spread metric NS@K decreases significantly as the 2nd-IF incorporates the spread capability of the candidates into consideration.

Table 3: Ablation study about the 1st-IF and 2nd-IF of the HeteroIR on Twitter dataset.

| Ablation | R@1 | R@2 | N@1 | N@2 | NS@1 | NS@2 |
|---|---|---|---|---|---|---|
| w/o 1st-IF | 0.2503 | 0.3837 | 0.2933 | 0.3540 | 0.2342 | 0.3298 |
| w/o 2nd-IF | 0.2849 | 0.4070 | 0.3302 | 0.3816 | 0.2258 | 0.3166 |
| HeteroIR | **0.2933** | **0.4244** | **0.3402** | **0.3969** | **0.2432** | **0.3425** |

**HeteroIM.** In HeteroIM, we utilize heterogeneous spread probability (HSP) and introduce a uniform sampling (US) strategy in RR set generation. Further, we rerank the candidates based on the shared RR sets (SRR). The results show that the rerank significantly improves the performance. As traditional IM methods [4, 33, 40] mainly focus on reranking based solely on the spread influence, regardless of the interaction willingness. Moreover, we find that incorporating the HSP and US into the generation of RR-sets improves the performance of our algorithm, as they introduce interaction willingness and lower the sampling bias in the sampling stage, respectively.

Table 4: Ablation study about HeteroIM on Twitter dataset.

| Ablation | R@1 | R@2 | N@1 | N@2 | NS@1 | NS@2 |
|---|---|---|---|---|---|---|
| w/o Rerank | 0.1176 | 0.2342 | 0.1416 | 0.2030 | 0.1845 | 0.2843 |
| w/o HSP | 0.3487 | 0.4643 | 0.4082 | 0.4475 | 0.2247 | 0.3199 |
| w/o US | 0.3598 | 0.4731 | 0.4221 | 0.4581 | 0.2439 | 0.3297 |
| HeteroIM | **0.3655** | **0.4997** | **0.4284** | **0.4780** | **0.2445** | **0.3543** |

## 4.4 Influence Algorithm Comparision (RQ3)

To validate the HeteroInf algorithm, we assess the performance of various algorithms in evaluating nodes' spreading influence. Within the game scenario, our objective is to identify a subset of highly influential users from the entire pool of users, encouraging them to invite their friends to participate in the activity. Consequently, we use Hit@K to evaluate the effectiveness of different algorithms in finding the influential users.

Hit@K quantifies the percentage of top-k selected users whose actual propagation ability ranks within the true top-k, reflecting the algorithm's capability to identify high-influence users. As shown in Figure 4, HeteroInf achieves the highest Hit@K values across varying K values while maintaining robust performance on different datasets. Notably, its superior performance at smaller K values demonstrates strong capability in selecting influential spreaders.

Despite node selection, we further perform a comparison on the recommendation task. The ranking function for the HeteroInf is described as $P_{uv}U_v + P_{uv}U_v I_H(v)$ while the $I_H(v)$ denotes the influence estimated by HeteroInf. Here, we compare different spread influence algorithms by shifting $I_H(u)$ to other algorithms, such as Monte Carlo influence $I_{MC}(u)$.

Figure 5 shows the comparison of the Linear Regression Influence [9], the Monte Carlo Influence [19], and the HeteroInf. We notice that the spread influence from the HeteroInf performed best among NSpread@K and NDCG@K, which is consistent with the performance in spread influence estimation.

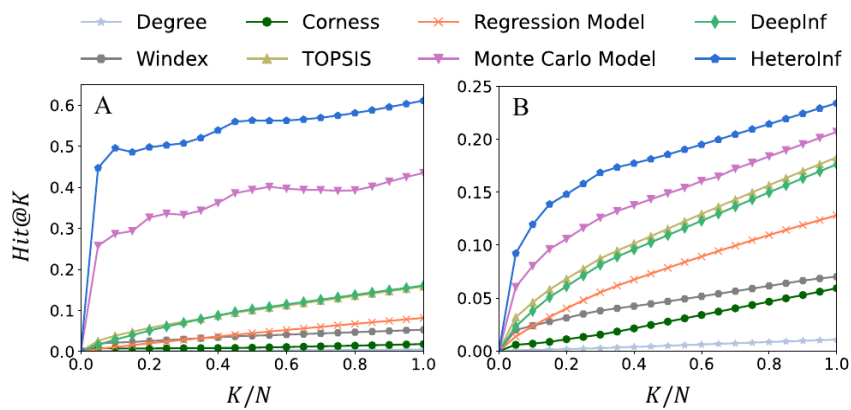

Figure 4: The performance of various influence algorithms on Hit@K in the TXG-B and C datasets.

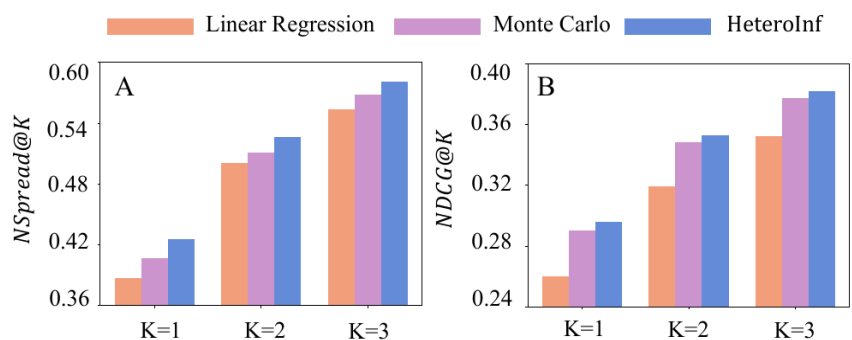

Figure 5: Comparison among spread influence algorithms in recommendation tasks from LR, MC, and HeteroInf algorithm on TXG-B dataset.

### 4.5 DEPLOYMENT (RQ4)

We deployed HeteroIR and HeteroIM on two propagation events in Tenctnt's first-person shooting game, referred to as X1 and X2. The treatment group of X1 and X2 contains 15.6 and 16.4 million users.

We implement HeteroIR and HeteroIM based on the invite probability $P_{uv}$ and the accept probability $U_v$ predicted. Then aggregate the HeteroInf with interact capability $w = 4$, which corresponds to the average spread capability in the game platform. We follow [30] and partition users into communities with high connectivity and profile homophily. We then conduct the online A/B test by randomly assigning users in the same communities to the same treatment group. We evaluated the performance based on three indicators:

(i) Secondary Invite Rate (Sec-IR) : Sec-IR describes the ratio of invitees invited with secondary spread.

(ii) Secondary Invite Times (Sec-IT) : Sec-IT describes the average secondary spread times of the invitees.

(iii) Reach Retain Rate (RRR) : RRR describes the ratio of invitees invited who log in to the game the next day.

Table 5 shows the A/B test performance on Event X1 and X2. The results demonstrate that HeteroIR and HeteroIM perform best among all the events. Specifically, HeteroIM gained the relative improvement of 10%, 9.64%, and 14.83% for SIR, SIT, and RRR, respectively, compared with intimacy in event X2.

Table 5: Performance on Tencent's propagation event X1 and X2. The bold value indicates the best performance.

| Event | X1 | | | X2 | | |
|---|---|---|---|---|---|---|
| Model | Sec-IR (%) | Sec-IT | RRR (%) | Sec-IR (%) | Sec-IT | RRR (%) |
| Intimacy | 8.095 | 0.184 | 17.109 | 5.911 | 0.197 | 30.644 |
| HeteroIR | **11.029** | **0.249** | **25.018** | 6.414 | 0.210 | 35.160 |
| HeteroIM | - | - | - | **6.502** | **0.216** | **35.191** |

## 5 RELATED WORK

**Influence Maximization**. Identifying influential nodes that drive rapid and widespread propagation within social networks is of significant theoretical and practical importance. Intuitive structure methods such as Degree centrality [38], Coreness [21] are widely used to quantify the spread influence of each user. However, directly using these metrics to rerank might lead to a high overlap among users. To overcome such challenges, Influence-Maximization algorithms are proposed to select a set of nodes to maximize the spread coverage [4, 14, 15, 33, 34]. Borges [4] proposed RIS to select the nodes based on the RR sets iteratively. IMM [34] grounded on the martingale theory, estimates that the lower bound of the RR sets leads to significant improvement in running time. OPIM-C [33] introduces adaptive bound tightening using intermediate greedy selection results, enabling both high flexibility and superior offline performance. Zhang [40] proposed RR-OPIM+ to generate capacity-limited candidates for the spread maximization. Even though the IM methods above can select a set of nodes with a high spread potential. However, these methods did not consider interaction willingness, which might lead to a setback in satisfying the basic requirement of the recommendation, such as click rate.

**Recommendation Algorithm**. Recommendation algorithms have predominantly focused on modeling the interaction willingness between the user and the candidates. In user recommendation tasks, the closeness of the relationship can be estimated by the local structure. Graph-based random-walk algorithm, personalized PageRank (PPR) [24], was proposed to calculate the correlation between the users in the social networks. To extend the estimation more precisely, the supervised method [31, 35, 44] was proposed to directly model click-through rates with parameters $\theta$ to predict the likelihood $P(y_i = 1|x_i, \theta)$ that a user will interact with a particular candidate $x_i$, which achieve great performance on the click rate prediction. However, existing methods fail to consider the spread potential of the candidates, which leads to limitations in spread maximization.

## 6 CONCLUSION AND FUTURE WORK

To fully unleash the potential of user recommendation systems in information dissemination while preserving interaction willingness, we introduce HeteroIR, an influence-based algorithm tailored to optimize both criteria. Additionally, we propose HeteroInf for improved estimation of personalized spread influence. To further mitigate the spread overlap among candidates while guaranteeing interaction willingness, we present HeteroIM, an algorithm grounded in the influence maximization framework. Extensive experiments validate the superiority of our methods in both spread and recommendation tasks. Moreover, deploying HeteroIR and HeteroIM in in-game propagation events has yielded notable enhancements. As a future direction, we desire to expand the personalized interaction capacity in HeteroInf and explore more efficient algorithms to end-to-end model invitation and acceptance probabilities.

## Acknowledgments

X. W. was supported in part by GuangDong Basic and Applied Basic Research Foundation (Grant No. 2025A1515011389, 2023B0303000009), and in part by Chinese NSF (Grant No. 62327808).

## GenAI Usage Disclosure

No Generative AI is utilized in this work.

## References

[1] Ra' Almestarihi, AYAB Ahmad, R Frangieh, I Abu-AlSondos, K Nser, and Abdulkrim Ziani. 2024. Measuring the ROI of paid advertising campaigns in digital marketing and its effect on business profitability. *Uncertain Supply Chain Management* 12, 2 (2024), 1275–1284.

[2] Eytan Bakshy, Solomon Messing, and Lada A Adamic. 2015. Exposure to ideologically diverse news and opinion on Facebook. *Science* 348, 6239 (2015), 1130–1132.

[3] Eytan Bakshy, Itamar Rosenn, Cameron Marlow, and Lada Adamic. 2012. The role of social networks in information diffusion. In *Proceedings of the 21st international conference on World Wide Web*. 519–528.

[4] Christian Borgs, Michael Brautbar, Jennifer Chayes, and Brendan Lucier. 2014. Maximizing social influence in nearly optimal time. In *Proceedings of the twenty-fifth annual ACM-SIAM symposium on Discrete algorithms*. SIAM, 946–957.

[5] Robert B Cialdini. 2009. *Influence: Science and practice*. Vol. 4. Pearson education Boston.

[6] Israel Cohen, Yiteng Huang, Jingdong Chen, Jacob Benesty, Jacob Benesty, Jingdong Chen, Yiteng Huang, and Israel Cohen. 2009. Pearson correlation coefficient. *Noise reduction in speech processing* (2009), 1–4.

[7] Manlio De Domenico, Antonio Lima, Paul Mougel, and Mirco Musolesi. 2013. The anatomy of a scientific rumor. *Scientific reports* 3, 1 (2013), 2980.

[8] Liguo Fei and Yong Deng. 2017. A new method to identify influential nodes based on relative entropy. *Chaos, Solitons & Fractals* 104 (2017), 257–267.

[9] David A Freedman. 2009. *Statistical models: theory and practice*. cambridge university press.

[10] Jacob Goldenberg, Barak Libai, and Eitan Muller. 2001. Talk of the network: A complex systems look at the underlying process of word-of-mouth. *Marketing letters* 12 (2001), 211–223.

[11] Mark S Granovetter. 1973. The strength of weak ties. *American journal of sociology* 78, 6 (1973), 1360–1380.

[12] Huifeng Guo, Ruiming Tang, Yunming Ye, Zhenguo Li, and Xiuqiang He. 2017. DeepFM: a factorization-machine based neural network for CTR prediction. In *Proceedings of the 26th International Joint Conference on Artificial Intelligence*. 1725–1731.

[13] Liyi Guo, Junqi Jin, Haoqi Zhang, Zhenzhe Zheng, Zhiye Yang, Zhizhuang Xing, Fei Pan, Lvyin Niu, Fan Wu, Haiyang Xu, et al. 2021. We know what you want: An advertising strategy recommender system for online advertising. In *Proceedings of the 27th ACM SIGKDD Conference on Knowledge Discovery & Data Mining*. 2919–2927.

[14] Qintian Guo, Sibo Wang, Zhewei Wei, and Ming Chen. 2020. Influence maximization revisited: Efficient reverse reachable set generation with bound tightened. In *Proceedings of the 2020 ACM SIGMOD international conference on management of data*. 2167–2181.

[15] Qintian Guo, Sibo Wang, Zhewei Wei, Wenqing Lin, and Jing Tang. 2022. Influence maximization revisited: efficient sampling with bound tightened. *ACM Transactions on Database Systems (TODS)* 47, 3 (2022), 1–45.

[16] Pankaj Gupta, Ashish Goel, Jimmy Lin, Aneesh Sharma, Dong Wang, and Reza Zadeh. 2013. Wtf: The who to follow service at twitter. In *Proceedings of the 22nd international conference on World Wide Web*. 505–514.

[17] Lucas Hanke and Luiz Chaimowicz. 2017. A recommender system for hero line-ups in MOBA games. In *Proceedings of the AAAI Conference on Artificial Intelligence and Interactive Digital Entertainment*, Vol. 13. 43–49.

[18] Ruining He and Julian McAuley. 2016. Ups and downs: Modeling the visual evolution of fashion trends with one-class collaborative filtering. In *proceedings of the 25th international conference on world wide web*. 507–517.

[19] Yanqing Hu, Shenggong Ji, Yuliang Jin, Ling Feng, H Eugene Stanley, and Shlomo Havlin. 2018. Local structure can identify and quantify influential global spreaders in large scale social networks. *Proceedings of the National Academy of Sciences* 115, 29 (2018), 7468–7472.

[20] Shangrong Huang, Jian Zhang, Lei Wang, and Xian-Sheng Hua. 2015. Social friend recommendation based on multiple network correlation. *IEEE transactions on multimedia* 18, 2 (2015), 287–299.

[21] Maksim Kitsak, Lazaros K Gallos, Shlomo Havlin, Fredrik Liljeros, Lev Muchnik, H Eugene Stanley, and Hernán A Makse. 2010. Identification of influential spreaders in complex networks. *Nature physics* 6, 11 (2010), 888–893.

[22] Joseph A Konstan and John Riedl. 2012. Recommender systems: from algorithms to user experience. *User modeling and user-adapted interaction* 22 (2012), 101–123.

[23] Yehuda Koren, Robert Bell, and Chris Volinsky. 2009. Matrix factorization techniques for recommender systems. *Computer* 42, 8 (2009), 30–37.

[24] Wenqing Lin. 2019. Distributed algorithms for fully personalized pagerank on large graphs. In *The World Wide Web Conference*. 1084–1094.

[25] Wenqing Lin. 2021. Large-Scale Network Embedding in Apache Spark. In *Proceedings of the 27th ACM SIGKDD Conference on Knowledge Discovery and Data Mining*. ACM, 3271–3279.

[26] Wenqing Lin, Feng He, Faqiang Zhang, Xu Cheng, and Hongyun Cai. 2020. Initialization for Network Embedding: A Graph Partition Approach. In *Proceedings of the 13th ACM International Conference on Web Search and Data Mining*. ACM, 367–374.

[27] Bryan Perozzi, Rami Al-Rfou, and Steven Skiena. 2014. Deepwalk: Online learning of social representations. In *Proceedings of the 20th ACM SIGKDD international conference on Knowledge discovery and data mining*. 701–710.

[28] Jiezhong Qiu, Jian Tang, Hao Ma, Yuxiao Dong, Kuansan Wang, and Jie Tang. 2018. Deepinf: Social influence prediction with deep learning. In *Proceedings of the 24th ACM SIGKDD international conference on knowledge discovery & data mining*. 2110–2119.

[29] Fernando P Santos, Yphtach Lelkes, and Simon A Levin. 2021. Link recommendation algorithms and dynamics of polarization in online social networks. *Proceedings of the National Academy of Sciences* 118, 50 (2021), e2102141118.

[30] Martin Saveski, Jean Pouget-Abadie, Guillaume Saint-Jacques, Weitao Duan, Souvik Ghosh, Ya Xu, and Edoardo M Airoldi. 2017. Detecting network effects: Randomizing over randomized experiments. In *Proceedings of the 23rd ACM SIGKDD international conference on knowledge discovery and data mining*. 1027–1035.

[31] Weiping Song, Chence Shi, Zhiping Xiao, Zhijian Duan, Yewen Xu, Ming Zhang, and Jian Tang. 2019. Autoint: Automatic feature interaction learning via self-attentive neural networks. In *Proceedings of the 28th ACM international conference on information and knowledge management*. 1161–1170.

[32] Jessica Su, Aneesh Sharma, and Sharad Goel. 2016. The effect of recommendations on network structure. In *Proceedings of the 25th international conference on World Wide Web*. 1157–1167.

[33] Jing Tang, Xueyan Tang, Xiaokui Xiao, and Junsong Yuan. 2018. Online processing algorithms for influence maximization. In *Proceedings of the 2018 international conference on management of data*. 991–1005.

[34] Youze Tang, Yanchen Shi, and Xiaokui Xiao. 2015. Influence maximization in near-linear time: A martingale approach. In *Proceedings of the 2015 ACM SIGMOD international conference on management of data*. 1539–1554.

[35] Zhen Tian, Ting Bai, Wayne Xin Zhao, Ji-Rong Wen, and Zhao Cao. 2023. EulerNet: Adaptive Feature Interaction Learning via Euler's Formula for CTR Prediction. In *Proceedings of the 46th International ACM SIGIR Conference on Research and Development in Information Retrieval*. 1376–1385.

[36] Aaron Van den Oord, Sander Dieleman, and Benjamin Schrauwen. 2013. Deep content-based music recommendation. *Advances in neural information processing systems* 26 (2013).

[37] Qiwei Wang, Dandan Lin, Wenqing Lin, and Ziming Wu. 2025. FROG: Effective Friend Recommendation in Online Games via Modality-aware User Preferences. In *Proceedings of the 48th International ACM SIGIR Conference on Research and Development in Information Retrieval, SIGIR 2025, Padua, Italy, July 13-18, 2025*. ACM, 2822–2826.

[38] Stanley Wasserman and Katherine Faust. 1994. Social network analysis: Methods and applications. (1994).

[39] Qiang Wu. 2010. The w-index: A measure to assess scientific impact by focusing on widely cited papers. *Journal of the American Society for Information Science and Technology* 61, 3 (2010), 609–614.

[40] Shiqi Zhang, Yiqian Huang, Jiachen Sun, Wenqing Lin, Xiaokui Xiao, and Bo Tang. 2023. Capacity constrained influence maximization in social networks. In *Proceedings of the 29th ACM SIGKDD Conference on Knowledge Discovery and Data Mining*. 3376–3385.

[41] Shiqi Zhang, Jiachen Sun, Wenqing Lin, Xiaokui Xiao, Yiqian Huang, and Bo Tang. 2024. Information Diffusion Meets Invitation Mechanism. In *Companion Proceedings of the ACM on Web Conference 2024, WWW 2024, Singapore, Singapore, May 13-17, 2024*. ACM, 383–392.

[42] Xingyi Zhang, Shuliang Xu, Wenqing Lin, and Sibo Wang. 2023. Constrained social community recommendation. In *Proceedings of the 29th ACM SIGKDD conference on knowledge discovery and data mining*. 5586–5596.

[43] Guorui Zhou, Xiaoqiang Zhu, Chenru Song, Ying Fan, Han Zhu, Xiao Ma, Yanghui Yan, Junqi Jin, Han Li, and Kun Gai. 2018. Deep interest network for click-through rate prediction. In *Proceedings of the 24th ACM SIGKDD international conference on knowledge discovery & data mining*. 1059–1068.

[44] Jieming Zhu, Qinglin Jia, Guohao Cai, Quanyu Dai, Jingjie Li, Zhenhua Dong, Ruiming Tang, and Rui Zhang. 2023. Final: Factorized interaction layer for ctr prediction. In *Proceedings of the 46th International ACM SIGIR Conference on Research and Development in Information Retrieval*. 2006–2010.